# Superconductors described with CSM.

*Department of physics, Nanjing University of Information Science & Technology, Nanjing 210044, China*

The properties of 'known' superconductors can be explained with the correlations dominated superconducting mechanism (CSM). The correlations have the spin correlation, the charge correlation and the spin-charge correlation, and their strengths can be described by the related correlation lengths in their correlation functions. Our evaluation from many superconductors is that superconductivities occur if both the spin correlation and the charge correlation are stronger, and the calculation of a Hubbard model showed that the spin-charge correlation may govern superconductivities[1]. Afterwards, this mechanism has led a model which includes various superconductivities and magnetisms, and the relation between superconductivities and magnetisms can be understood on this model[2] (these results have been shown by calculations). This mechanism is very practical, for example, to turn a material into a superconductor or increase the $T_c$ of a superconductor, what we will do is to increase the spin-charge correlation. In this letter, we first describe the relations between the spin-charge correlation, the spin correlation and the charge correlation, take these relations as the basis of constructing a new phase diagram, and then classify the 'known' superconductors into various sections in this phase diagram. This letter also gives a new explanation about the pressure effect on $T_c$, the isotope effect on $T_c$ and the pairing symmetry with the CSM.

We consider the bulk electrons. It is certain that correlations are affected by various factors, such as impurities, disorders, temperature, external fields, pressure, material structures etc. A way to understand the superconducting mechanism is studying the relation between superconductivities and correlations, and then studying the relations between correlations and various factors. This is a common method in theory: if we understand the relation between A and C, and understand the relation between B and C, we should understand the relation between A and B. One knows that the charge correlation increases with the decreasing of the spin correlation, while both the spin correlation and the charge correlation are beneficial for the spin-charge correlation, thus I divide the phase diagram into the sections 1-6 as shown roughly in Fig.1.

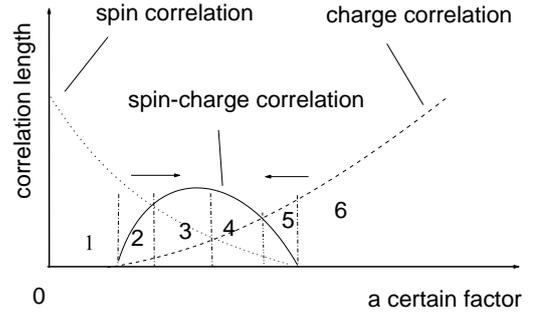

Figure 1: A schematic of the relation between three correlations.

One could give a clear schematic for the change of correlations, for example, the antiferromagnetic correlation length vs. doping in a definite temperature for $Nd_{2-x}Ce_xCuO_{4-y}$, or $La_{2-x}Sr_xCuO_{4-y}$ etc. A material has its special phase diagram, and their details should be more complex than the one in Fig.1. The charge correlation is neglected in the section 1 in which electrons is in insulating states, and such electron systems may appear magnetism when spin correlation is strong enough. The spin correlation is neglected in the region 6 where electrons are in the 'good metal states', the elements Au, Ag and Cu are such metals. The possible superconducting region is in the sections 2-5 where the spin-charge correlation is not zero. Of course, the superconducting state may be unstable in the section 2, phase separation might occur, but this is not our focus in this letter. It is seen that the more quickly the spin correlation changes with some factors, the more weak the spin-charge correlation is, and the more narrow the superconducting region is. $Nd_{2-x}Ce_xCuO_{4-y}$ may belong to this case, its superconductivity occur at a narrow doping region, but its section 1 is in a wider doping region. However, few correlation lengths have been exactly known for superconductors in theory. Then, we must find some understandable features to express these correlation strengths.

An interesting topic is the excitations what superconductivity originates in. The idea that superconductivity is dominated by the spin-charge correlation runs through this paper, but the spin-charge correlation is related to both the spin correlation and the charge correlation, thus the spin-charge correlation can be strengthened by the spin correlation or the charge correlation as shown in Fig.1. It is easy to understand that the spin excitations (the spin density wave or other magnetic fluctuations) appear when some spin correlation exists in a material, while the charge excitations or phonons appear when the charge correlation exists in a material. Therefore, an inevitable deduction is that the superconductivity should be governed by all excitations. There is shown by this appearance that superconductivity is resulted from disorder, magnetic field, or pressure etc in some known experiments. On the basis of these appearances, we argue that the spin excitations will dominate the superconductivity when the spin correlation is the strongest one, while the charge excitations or phonons will dominate the superconductivity when the charge correlation is the strongest one. One may question why the magnetic excitations of some materials will be increased when they are coming into superconducting states or increasing $T_c$ [3]. The CSM argues that increasing the spin-charge correlation requires increasing the spin correlation.

The possible excitations should be argued combining with the pairing symmetry, and this is the second topic. It is necessary to note that the short-length interaction is related to anisotropic properties, while the long-length interaction is related to isotropic ones. The long-length interactions are important for electron-phonon interaction, this interaction behaves as isotropic, and thus phonons mediated superconductivity is s-wave symmetry as argued in BCS. On the contrary, the short-length interactions are important for the spin mediated interaction, this interaction behaves as anisotropic in crystals, and thus spins mediated superconductivity has the d-wave symmetry or the anisotropic s-wave symmetry. On the basis of this idea, the charge excitations induced superconductivity could be the s-wave symmetry or the d-wave symmetry since the electron-electron interaction can be the long-range or short-range one. Thus we arrive at the tableⅠ:

tableⅠ: The low temperature features, pairing symmetry and excitations vs. the phase diagram.

| excitations | spins | charges | | phonons |
|---|---|---|---|---|
| Sections in Fig.1 | 2&3 | 4 near3 | 4 near 5 | 5 |
| Pairing symmetry | d-wave | d--wave | s-wave | s-wave |
| behavior | Non-FL | Non-FL | FL | FL |

One may believe that the electron system behave as the non-Fermi liquid (Non-FL) when the spin excitations mediate the pairing and the electron systems behave as the Fermi liquid (FL) when phonons mediate the pairing, but a possible debate may be whether these electrons behave as the FL when charge excitations dominate physics, and this is one of my propositions. The reason is that the electrons appear the spin-charge separation (the short-length interaction is related to the strongly correlated interaction.), the electrons should behave as the non-FL, and the charges induced superconductivity is the d-wave symmetry. Otherwise, the electrons appear charge excitations in other ways, the charges mediated superconductivity should be the s-wave symmetry.

That is to say, on the basis of the tableⅠ, if we know the FL or the non-FL behavior and the pairing symmetry of electron systems, we can distinguish whether the superconductivity is mediated by spin excitations, charge excitations or phonons. Therefore, we conclude: (1) BCS superconductors are induced by phonons. (2) The MgB2 superconductor is dominated by charge excitations due to their properties. (3) The underdoped and optimally doped p-type cuprate superconductors are dominated by spin excitations, while the overdoped and electron-doped cuprates and heavy Fermions superconductors are dominated by charges. (4) Some of FeAs-based superconducting compounds and heavy fermion superconductors can be clarified, and others have to be judged in future because few of their properties are observed in experiments to my knowledge. In a word, various 'known' superconductors could be described with this CSM. These results extend the tableⅠ to the tableⅡ:

tableⅡ: Superconductors vs. CSM

| excitations | spins | charges | phonons |
|---|---|---|---|
| Sections in Fig.1 | 2&3 | 4 near3 | 4 near 5 | 5 |
| Pairing symmetry | d-wave | s-wave | |
| behavior | Non-FL | FL | |
| Super-conductors | under & optimally doped p-type cuprates; $K_{0.2}Sr_{0.8}Fe_2As_2$ | BCS ones; $MgB_2$; n-type cuperates; overdoped cuprates; heavy fermion ones; MgCxNi3 | |

The third topic is the isotope effect. The isotope effect on $T_c$ is related to the interactions between electrons. For example, if the d-holes are more important than p-holes for underdoped p-type cuprates where the short-length interaction dominates the superconductivity, the substitution effect of Cu isotope element should be evident. On the contrary, the O substitution is evident for overdoped p-type cuprates. For example, that the $Zn^{2+}$ substitutes $Cu^{2+}$ in $CuO_2$ planes[4] evidently affect the $T_c$ is because both the spin correlation and the charge correlation are evidently weakened by this substitution, and the spin-charge correlation is evidently weakened when both the charge correlation and the spin correlation are not strong enough as shown in Fig.1. For example again, the Tc of Al is 1.2K and the Tc of Nb is 9.3K, while the Tc of $Nb_3Al$ is 17.5K, why? The charge correlation in Al material is stronger, while the spin correlation in Nb is not weak enough, and then the spin-charge correlation in $Nb_3Al$ is strengthened compared with the one in ether Cu or Nb. The $Nb_3Al$ is a metal compound, but we also say 'Al substitutes some of Nb' despite this word is not very right.

The next topic is the physical pressure effect on superconductivity. There are many explanations about this effect in literatures, while we suggest that the Tc of a material can be raised with increased pressure as soon as the spin-charge correlation of the material is strengthened. Many FL compounds[5,6] showing a negative pressure coefficient of $T_c$ because they belong to the 'charge correlation governed region' as shown in Fig.1, the increased pressure usually increase the charge correlation and decrease the spin correlation so that the spin-charge correlation is weakened because the carrier concentration is increased. However, the $T_c$ of underdoped or optimally doped p-type cuprate superconductors can be increased with the increased pressure in some pressure range[7,8,9] because their spin-charge correlations are strengthened through the increasing of the charge correlation. It is necessary to note that the spin-charge correlation is related to both the spin correlation and the charge correlation, thus the spin-charge correlation can be strengthened by the spin correlation or the charge correlation. As shown in Fig.1, the pressure will increase the $T_c$ of superconductor following the direction of increasing spin-charge correlation shown by arrows →&←. On the contrary, the $T_c$ will decrease.

Let us now present some ways to confirm the CSM. For example, experiments[10,11,12,13] show that $MgC_xNi_3$ have the s-wave superconductivity, and $dT_c/dp$ >0. On the basis of the CSM, this material is in the section 5 or 4 near 5, the pressure should increase the spin correlation of electrons as shown in tableⅡ, thus non-FL behavior should appear with the increasing of the pressure as soon as the material comes into the section 3, and this prediction has to be confirmed. Another example is $K_{0.2}Sr_{0.8}Fe_2As_2$, authors reported that $dT_c/dp$ >0 , $d\rho/dp$ <0 and the SDW transition temperature $T_S$ (not shown in the figure) decreases from 139 K (p=0) to 129 K at 15 kbar[14]. I suggest that $K_{0.2}Sr_{0.8}Fe_2As_2$ is in the section 2 or 3 in Fig,1, thus its superconductivity should be d-wave symmetry with the CSM, and this has to be confirmed by experiments.